\documentclass{sigchi}

\usepackage{balance}  
\usepackage{graphics} 

\usepackage{txfonts}
\usepackage{times}    
\usepackage[pdftex]{hyperref}
\usepackage{color}
\usepackage{textcomp}
\usepackage{booktabs}
\usepackage{ccicons}
\usepackage{todonotes}
\usepackage{changepage}
\usepackage{amsmath,amssymb}
\usepackage{url}
\usepackage{array}

\makeatletter
\def\url@leostyle{%
  \@ifundefined{selectfont}{\def\UrlFont{\sf}}{\def\UrlFont{\small\bf\ttfamily}}}
\makeatother
\def\pprw{8.5in}
\def\pprh{11in}

\definecolor{linkColor}{RGB}{6,125,233}

\definecolor{jess}{RGB}{102,166,30}
\definecolor{jomara}{RGB}{240,128,128}
\definecolor{ben}{RGB}{72,118,255}
\definecolor{jack}{RGB}{217,95,2}

\hypersetup{%
  pdftitle={SIGCHI Conference Proceedings Format},
  pdfauthor={LaTeX},
  pdfkeywords={SIGCHI, proceedings, archival format},
  bookmarksnumbered,
  pdfstartview={FitH},
  colorlinks,
  citecolor=black,
  filecolor=black,
  linkcolor=black,
  urlcolor=linkColor,
  breaklinks=true,
}

\begin{document}

\title{Using Key Player Analysis as a Method for Examining the Role of Community Animators in Technology Adoption}

\numberofauthors{3}
\author{%
  \alignauthor{Jomara Sandbulte\\
    \affaddr{Pennsylvania State University}\\
    \affaddr{University Park, PA USA}\\
    \email{jmb89@psu.edu}}\\
  \alignauthor{Jessica Kropczynski\\
    \affaddr{University of Cincinnati}\\
    \affaddr{Cincinnati, OH USA}\\
    \email{jess.kropczynski@uc.edu}}\\
  \alignauthor{John M. Carroll\\
    \affaddr{Pennsylvania State University}\\
    \affaddr{University Park, PA USA}\\
    \email{jmc56@psu.edu}}\\
}

\maketitle


\section{Abstract}

This paper examines the role of community animators in technology adoption. Community animators are individuals that actively build social networks and broker ties between nodes in those networks. The present study observes technology adoption patterns through data collected from a mobile application at a local arts festival. A social network was constructed through photo-sharing and interaction within the app. Given this data, we propose the use of key player analysis to identify community animators. In addition, we use a graph invariant (i.e., fragmentation in the network) to describe the role and impact of key players on the full network of interactions. Our results contribute to literature on technology adoption in usability studies by proposing a method to quantify and identify the theoretical concept of community animators. We further analyze the types of community animators to be found in early adoption of technology: the early adopters themselves, and the initiating developers.



\section{Introduction}
In the digital age, much of our daily life is experienced using mobile devices and social media. There has been a growing interest in leveraging the benefits of mobile technology in many local communities to enhance cooperation among citizens and promote collective action \cite{hansen2014civic,ludwig2016publics,niederer2016smart,schafer2016challenging}. Prompting interactions among citizens has the potential to positively impact the economy of local organizations and markets \cite{campbell2013beyond,enacademic2015,han2016being}. Community events are important to economic development and regional tourism, but also build community pride. According to Wellman \cite{wellman1999network}, there are a few central goals of a local community. These consist of delivering local information to people, increasing peoples' awareness and participation in local activities, and bringing people closer to the local community. 

Although the goals and ascribed traits of communities may appear simple in nature, they can contain a great deal of nuance and complexity \cite{carroll2014neighborhood}. As such, designing for these goals and achieving critical mass in the study of community interaction within socio-technical systems can be a daunting task. Research has shown that individual psychology and personal influence are key factors in technology adoption \cite{lu2005personal}. Identifying and employing community animators may ease the task of reaching critical mass in socio-technical work. The present study examines both the individuals who seed a network with interaction, and those who emerge as community animators themselves.

The interactions in this study are situated in the context of an arts festival promoted by a local community. As a community event established in 1967, the festival has become an annual tradition for many local families, students, and visiting alumni of the local university. Annual attendance is estimated to be almost 150,000 over the lifespan of the festival. The size of the event, coupled with wide usage of smart phones and other mobile devices, presents unique opportunities for documenting and researching networked interactions among festival-goers. Twitter posts, photo tagging, geolocation information, and myriad other components of social media create opportunities for community engagement. Mobile applications are able to promote interpersonal interactions by providing community members with new modes of communication and experience sharing \cite{ganoe2010mobile,han2014local}. In these networks, mobile technology seeks to improve the quality of interaction between people. This increase in quality may lead to a heightened sense of community engagement \cite{mcmillan1986sense}.

Promoting positive interactions in local events may be an effective method for strengthening communities. However, these interactions do not typically occur without some form of incitement or prompting. Social intermediaries work to facilitate such positive interactions by building bridges within the community between disparate network components. This act is essential in promoting technology adoption. We conceptualize this as the role of a Community Animator. Community animator is an emerging job title within innovation spaces, such as incubators and maker environments \cite{bjorgvinsson2010participatory,coalition2015,linkedin2015,shih2015engaging}. These animators participate in and manage communities by sharing information and building bridges between members and groups \cite{animation2010,knight2010,linkedin2015}. Community informatics research has examined the role of bridges, particularly in the form of community leaders, as a way to integrate communities\cite{carroll2003blacksburg}. In our research, we view the socio-technical system as a tool to enhance the work of
community animators and facilitate bridging among those
that do not frequently take on this role.

Many online social networks begin with interactions among
developers, or a select group of early adopters. These early
adopters often become animators of the space. For example,
Mark Zuckerberg, who started Facebook in his college dorm
room \cite{phillips2007brief}, and his friends were the initial users of the social
network and seeded initial interactions. We recognize the
value of initial interactions and understand that the success
of many apps can be attributed to these early steps.

In our study, we consider the technology developers and
research staff that work to gain initial usage as seeded
animators. The interactions they create serve to promote
shared activity among the festival's participants and
encourage the creation of new community animators. Our
research aims to investigate the impact of seeded animators
on a network of socio-technical interactions. To better
understand the impact of community animators' on
technology adoption, we begin by proposing key player
analysis to identify and quantify the role.

In social network analysis, degree of centrality (or the node
with the most interactions) is often used to describe nodes of
some relative importance to a network or a node with higher
degrees of information flow to a particular node \cite{borgatti2005centrality}. Borgatti \cite{borgatti2006identifying}
described the tendency to measure importance of nodes
with centrality as part of a key player problem and developed
key player analysis and corresponding software to
distinguish between popular nodes (with a high degree of
centrality) and nodes that have other structurally
advantageous positions. Borgatti's key player analysis
software utilizes a set of algorithms to identify sets of nodes
that 1) if removed, fragment or disrupt the network, and 2)
are able to reach the largest proportion of the network.

We present two socio-technical systems for encouraging
interactions and utilize key player analysis to identify
community animators’ role in system adoption. First, we
designed a social game that blends the use of mobile devices
with a social infrastructure for face-to-face community
interactions. Second, we designed a mobile application that
contains a digital program and promotes lightweight social
interactions. We will describe how each social activity
encouraged interactions at the festival. Further, we will
discuss how key player analysis of seeded animators and
festival-goers can be used to identify individuals of
importance to the desired interactions. Finally, we present
our findings and relate them to lessons learned about the role
of community animators, and conclude suggestions for
future work.

\subsection{Research Questions}
Gaining critical mass in socio-technical systems can be
challenging, the initial work of developers and early adopters
are often critical to animating a socio-technical system and
ensuring its success. To better understand this role, we
borrow key player analysis \cite{borgatti2005centrality} as a method to articulate and
operationalize the conceptual definition of the community
animator. Our work is guided by the overarching question:

\textit{(RQ1) How can key player analysis be used as a method for
identifying community animators?}

Further, we ask: 

\textit{(RQ2) How do the networked interactions of seeded
developers differ from those of early adopters?}

\section{Related Work}
We use the term Community Animators to describe the role
of individuals that host many introductions in our sociotechnical
system. This section describes the emergence of
community animators and their role in community hubs.

\subsection{The Emergence of Community Animators}
Over a several years working with community partners in a
local community, \cite{anonymous}, it became clear to us that the collective
vision of local businesses, organizations, educational
institutions, and local government are often disconnected and
can become unaware of common goals for positive change.
Their activities were often stuck within silos defined by
generation, discipline, or interest levels. Reversing this effect
can reduce uncertainty, miscommunication, and misdirection
in smart and connected cities \cite{campbell2013beyond}. 

Community animators have been defined as ``a person who
works with the members of a community to help get them
informed and excited about local issues" \cite{enacademic2015}. A common
title for individuals that fill this structural role in a social
network is brokers, as introduced by Burt \cite{burt2005brokerage}. Brokers are
intermediary links in systems of social, economic, or
political relations who facilitate trade or transmission of
valued resource that would otherwise be very difficult.
According to Burt, the crucial characteristics of brokers are
that they (i) bridge gaps in social structure and (ii) help
goods, information, opportunities, or knowledge to flow
across those gaps\cite{burt2009structural}. The role of broker has much value to
the community in terms of strength in community vision and
having collective action taking place for example a broker
would take action in a local food bank to find people to
donate or to value the food bank \cite{burt2009structural, stovel2011stabilizing}. The modern
definition of a community animator in incubators and
innovation spaces extends beyond the traditional notion of a
brokers in sociology to include not only bridging, but also
taking action on needs when appropriate \cite{coalition2015}. Similar to
bridges, community animators work to connect members of
the community and facilitate knowledge, which may cause
high workload and less resilience.

In this study, we operationalize the community animator in
technology adoption as individuals that aid in building
cohesive networks of interactions among participations in a
socio-technical system. While interacting in an online
environment is a low-level interaction to ``animate"
communities, the role of helping other become informed
about a type of technology and helping them to be excited
about using it closely aligns with the description of this role
in face-to-face scenarios. In addition to inspiring participation, we hypothesize that community animators can be detected using algorithms that detect individuals that build cohesion among subgroups and clusters of participants that would otherwise be disconnected components of the network.

\subsection{Promoting Community Engagement}
Carroll \cite{carroll2015community} describes the community informatics as a field concerned with the challenges and opportunities for citizens in an environment increasingly dominated by technology. Instead of encouraging social interaction with people around local places, researchers \cite{Memarovic:2012:DIP:2406367.2406420,turkle2012alone} have affirmed that the use of information and communication technologies (ICT) also holds the potential to isolate society and motivate citizens to remain focused on their own isolated tasks. Ogilvy and Mather have responded to this shift by creating a series of public service announcements in China called ``The Phone Wall" to illustrate how technology is effecting family relationship and used the caption ``The more you connect, the less you connect" at the bottom of the images \cite{yudin2015}. Even though this seems to be the primary impact of ICTs, they have also contributed to the development of applications to support local economy \cite{campbell2013beyond}, social justice \cite{starbird2012will}, political empowerment \cite{campbell2013beyond,patel2013emergence}, and collaboration and social exchange by using social networking technology \cite{jarrahi2012social}.

Considerable efforts have been dedicated to the development of applications that integrate innovations and technology to facilitate community citizen engagement. Researchers have studied citizens' engagement in local community events such as festivals. Cheverst et al. \cite{cheverst2008supporting} have employed applications for festivals to investigate social support through user generated mash ups of geo-tagged photos. Shih et al. \cite{shih2015engaging} have explored how social media discussions and photos content created during community event can directly enhance festival experience. In addition, Han et al. \cite{han2016being} explored communities conceptual framework to investigate how using and interacting with festival app enhance user's festival experience. Researchers in CSCW have also developed designs to inspire face-to-face meet-ups and engage citizens in their communities. Hansen et al. \cite{hansen2014civic} explored socio-technical platform to promote civic action brokering in local community, and Agarwala's \cite{agarwala2013researchbroker}study focused on a community of academic researchers developing an online system to match researchers with organizations or companies opportunities.

Researchers have also focused on context aware computing within community networks. Burrell \cite{burrell2002graffiti} has used context aware computer and mobile applications to virtually connect users in established communities. Ganoe et al. \cite{ganoe2010mobile} apply a location-sensitive mobile application for community engagement. However there is a dearth of research focusing on strategies to encourage a critical mass of participants to research these systems by creating local networks of active participants. Our research contributes this gap by analyzing our own efforts to inspire citizen engagement.

\subsection{Early Adoption and Technology Diffusion}
Research has strived to understand factors affecting software adoption \cite{shih2015engaging}. Developers argue that functionality is not the key driver to successful technology adoption \cite{enacademic2015} —in many cases, having a critical mass is crucial for long term sustainability \cite{linkedin2015,ludwig2016publics, shih2015engaging, whittle2010voiceyourview}. These studies have shown that critical mass is typically gained through techniques that are effective at recruiting early adopters. Early adopters and the people they affect allow for the rapid diffusion of technology and software \cite{droge2010lead} . The need to establish a base of early adopters is so critical that many developers have released their products completely free for a limited time \cite{whittle2010voiceyourview}. 

Of equal importance is the environment and state in which a particular technology is released to consumers. In addition to information content \cite{schafer2016challenging}, Merkel et al. \cite{merkel2007managing} examined the design and use of technology at nonprofit organizations and found community and organizational network structure to be crucial for technology adoption. These results indicate a need for understanding communities and demographics on both a structural and content-specific level. This regularly inspires citizen engagement projects to construct strategies for increased adoption in order to improve the robustness of future analyses \cite{jarrahi2012social}.

Early adopters are often critical to marketing strategies, and have been documented in previous studies to not only participate themselves, but invite others \cite{turkle2012alone}. Their engagement and influence on technology diffusion has also been seen in the communications of inter-organizational networks \cite{czepiel1975patterns}. Due to the influence-driven and community-oriented nature of ICTs' early adopters, the present study classifies them as community animators \cite{hansen2014civic}.

We are interested in examining the community animator's role in initiating interaction among local festival participants. In quantifying the value of seeded interactions to technology adoption, insights into networks of socio-technical systems can be gained. The present study aims to test the following hypotheses:

\begin{enumerate}
    \item Community Animators can be identified using Key Player Analysis
    \item Community Animators are important as ``bridges" connections
    \item Community Animators aid in technology adoption
\end{enumerate}

\section{Methods}
Since 2008, our team has developed the official mobile application for a large, annual arts festival (ArtsFest) that has a long tradition in the local community. In previous prototypes, the ArtsFest mobile application included an interactive calendar of events, but the ArtsFest 2014 application was the first version to contain social interactive features. At the 2014 ArtsFest, we explored how social infrastructures aided by mobile technology can increase social connections and sense of community. For the purposes of this study, we focus heavily on the social interactions
portion of the study. Different social activities were explored 2014 and 2015 during the festival. In 2014, we asked participants to engage in a selfie game and post selfies with other festival-goers in the festival app. In 2015, lightweight in-app social interactions were implemented such as waves to other festival-goers based on similar interest, photo-liking, and tagging. Based on in-app activities each year, we constructed an adjacency matrix of app interactions to construct a social network.

In addition to developing the applications, members of our lab staffed a booth at the festival to promote the mobile application and offer technical support.

\subsection{Selfie Game}
Festival-goers downloaded the ArtsFest 2014 application on both iOS (n=1025) and Android (n=413) platforms. A subsample of participants agreed to participate in a social game (n=150) intended to encourage content creation and to increase the sense of community among app users. The Selfie Game proposal is described in Figure 1\label{selfiegame}.

\begin{figure}[!ht]
\centering 
\includegraphics[width=9cm]{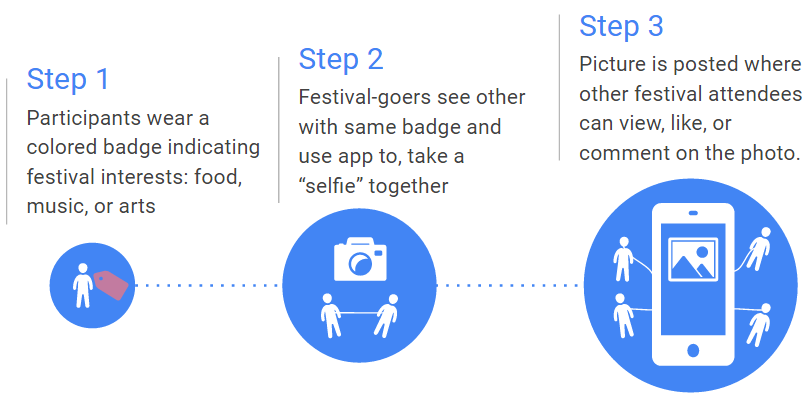}
\caption{Selfie Game: How it works}
\label{figure:selfiegame}
\end{figure}

Figure 2 illustrates festival-goers participating in the Selfie Game (Figure 2, left) and posting the “selfie” photo (Figure 2, right).

\begin{figure}[!ht]
\centering 
\includegraphics[width=9cm]{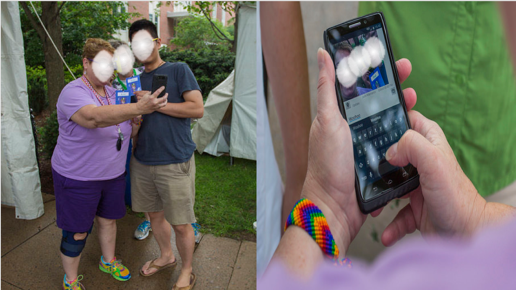}
\caption{Festival-goer participating in the Selfie Game}
\label{figure:selfiegame_participants}
\end{figure}

To create a network of interactions to utilize in our analysis, photos in the app were used to identify festival-goers that participated in the selfie game. Each festival-goer that uploaded a selfie constituted a node in the social network, the nodes were linked by appearing in a selfie together. If three individuals appeared in a selfie together, they would appear as a fully connected triad in the network. If one member of the triad then took a selfie with a passerby, the new node would become a pendent linked to triad by the individual that appeared in two selfies. Interactions were recorded and formatted into a person-by-person adjacency matrix.

\subsection{ArtsFest App}
In 2015, the ArtsFest social features were available on iOS (n=980). Features were designed based on the cooperation between our team lab and the festival administration. The festival administration contributed digital content such as event descriptions, images, and links to artists. Our development team organized the data in our database server and wrote APIs to enable communication between mobile clients and the server. The application activity data that was archived includes ``waves" from festival-goers to other festival-goers based on recommendations; photo sharing; commenting on pictures; liking pictures; a ``leaderboard" of top users; RSVPing to events; and a map of all pictures posted. Analysis of activity data compared participants in the social game that went on to contribute to content creation in the app (by posting photos and selfies) to those that did not. Interactions stemming from one festival-goer to another in the app was recorded as an interaction. The interactions among festival-goers was used to create an adjacency matrix to be analyzed as a social network.

\subsection{Data Analysis}
In analyzing the two social networks, we state two main goals:

\begin{enumerate}
    \item Key Player Analysis to identify Community Animators
    \item Distinguish between roles and positions of Seeded Developers and Early Adopters
\end{enumerate}

In understanding how key player analysis can be used to identify community animators, we explore two methods of key players analysis introduced by Borgatti \cite{borgatti2006identifying}. The first type of key player analysis described by Borgatti as key player problem/negative (KPP-NEG). Actors whose removal disrupts the network (by dividing it into multiple components) often play the role of bridges within a network. These actors are the social glue that hold segments of the network together. A graph invariant, or whole network measure to describe a characteristic of a network, can be used describe the impact of a node if removed. In this case, fragmentation of the network is used to describe the impact of the removal of key players on the overall cohesion of the network. Fragmentation describes the proportion of pairs in the network that cannot reach each other in the network. Fragmentation scores vary from zero to one, a score of one would indicate complete fragmentation (a network of isolated nodes) and score of zero would indicate a maximally connected network.

The second type of key player analysis described is called key player problem/positive (KPP-POS). Actors with the largest reach in the network are in a position to optimally diffuse information through a network. This procedure identifies a set of nodes that are positioned in the network in
a way that allows the largest proportion of the network to be reached. We contrast this to key players that may act as bridges in the network of interactions and consider the structural advantages of community animators that occupy each of these roles in the discussion section. Key player positive provides a descriptor of the percentage of unique nodes reach by the set of nodes identified. KPP-POS was used to help determine the number of nodes to select in each network by aiming to reach over 90\%\ of the network.

We used KeyPlayer \cite{borgatti2005centrality}software to find sets of key players in the network, both KPP-NEG and KPP-POS and note actors affiliation with the development team. Based on the results of KPP-POS, size of network, and number of seeded developers that participated in each social activity, we varied the size of the set of key player to identify in each network. In the 2014 selfie game, two members of the development team were active, in this network; five key players were used in each procedure. In the 2015 in-app activities, six members of the development team were active and ten key players were used in each procedure. After identifying the key nodes, we used UCINET 6 software for Windows to verify the degree of fragmentation in the network. In addition, to generate the images, we used NetDrawand NodeXL network visualization tools.

\section{Results}
\subsection{Community Animators in the Selfie Game}
While our analysis of the social structure of interactions is the focus of the analysis, some differences existed in terms of level participation among the overall group of 150 selfie game participants. Selfie game participants posted 142 photos to the app, of those photos, 125 photos were selfies. In the social network of selfie game interactions, 65 people appeared in selfies, comprising a network of people appearing in selfies together. The maximum number of people in a connected component within this network was 38. Figure 3 shows the Selfie Game sociogram of interactions.

\begin{figure}[!ht]
\centering 
\includegraphics[width=9cm]{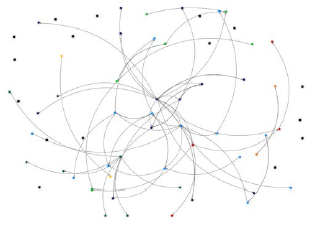}
\caption{Selfie Game Sociogram}
\label{figure:sociogram}
\end{figure}

After first visualizing this sociogram, it was immediately apparent that one of our active developers was a central actor in the network as is evident when highlighting his interactions within the network. Given his commitment to the project, we can affirm he had significant impact promoting interaction among Art Fest’s visitors as illustrated in Figure 4.

\begin{figure}[!ht]
\centering 
\includegraphics[width=9cm]{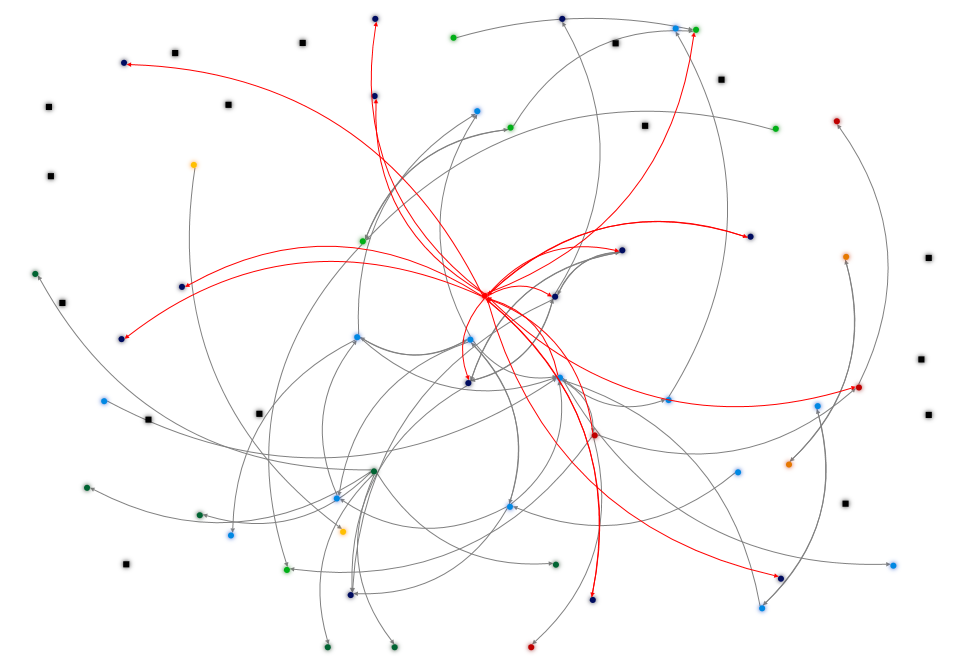}
\caption{Selfie Game Sociogram - with most active Community Animator highlighted}
\label{figure:sociogram_active}
\end{figure}

The question remained, as to how key player analysis can be used to identify actors in this position (RQ1). When performing KPP-NEG five key players were identified that held potential to fragment 98\%\ of the network. KPP-POS identified five key players that reached 93\%\ of the network. Of the actors in the network with the highest degree of centrality, the individual with the highest degree of centrality was an early adopter, and second highest was a seeded developer (same as highlighted in Figure 4) who obtained many interactions while staffing the booth at the festival, and the next three were early adopters of the app. A comparison of KPP-NEG, KPP-POS, and actors with the highest degree centrality are displayed in Table 1 in terms of the distribution of seeded developers and early adopters. Further examination of the impact of community animators is described in an analysis of fragmentation based on KPP-NEG in the proceeding section.

\begin{table}[!ht]
\setlength\extrarowheight{3pt}
\centering
\caption{Seeded Developers and Early Adopters Identified.}
\label{table:demographics}
\begin{tabular}{p{0.2\textwidth}|p{0.04\textwidth}|p{.04\textwidth}|p{.09\textwidth}}
\hline
\textbf{} & \textbf{KPP-NEG} & \textbf{KPP-POS} & \textbf{Highest Degree Centrality}\\
\hline
Seeded Developers & 2  & 1 & 2 \\ 
\hline
Early Adopters & 3  & 4 & 3 \\ 
\hline
\end{tabular}
\end{table}

Consistent with our first hypothesis, KPP-NEG effectively identified the two seeded developers (including the active developer in Figure 2). In addition, three early adopters were identified. The three early adopters identified using KPP-NEG were not the same individuals identified using KPP-POS.

To further understand the impact of community animators in the network, examination of a graph invariant, specifically
fragmentation, is used to characterize the cohesion that occurs through the presence of seeded developers and early adopters (RQ2). Before removing animators identified using KPP-NEG, the initial fragmentation in the network was 0.854. Figure 5 displays the full network with nodes sized by the degree of centrality. The nodes colored in orange represent seeded developers, green represent festival-goers who are early adopters, and nodes in blue were not identified as community animators (Figure 5).

\begin{figure}[!ht]
\centering 
\includegraphics[width=9cm]{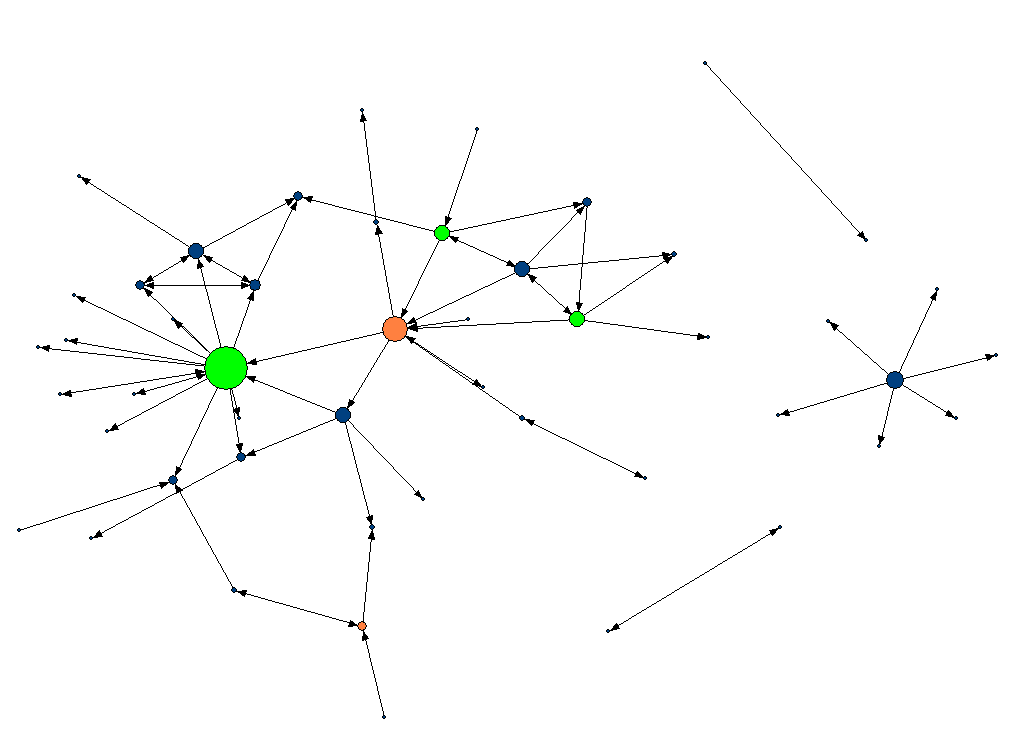}
\caption{Selfie Game with KPP-NEG highlighted (Orange= Seeded Developers Green=Early Adopters; Nodes sized by Degree Centrality).}
\label{figure:sociogram_kpp_neg}
\end{figure}

After removing seeded developers (Figure 6) fragmentation increased from initial fragmentation to 0.947, which represents a change in fragmentation of 0.093 (see Table 2). When early adopters were removed (retaining seeded developers as shown in Figure 6), the fragmentation of the remaining network is 0.895, representing a change of 0.041 in the fragmentation (see Table 2). Although only two of five community animators were seeded developers, their impact on the network is double that of the three early adopters. The difference between the two groups, but in bridging interactions among connected components, we can affirm that developers played an important role to promote interaction among festival attendees.

\begin{figure}[!ht]
\centering 
\includegraphics[width=9cm]{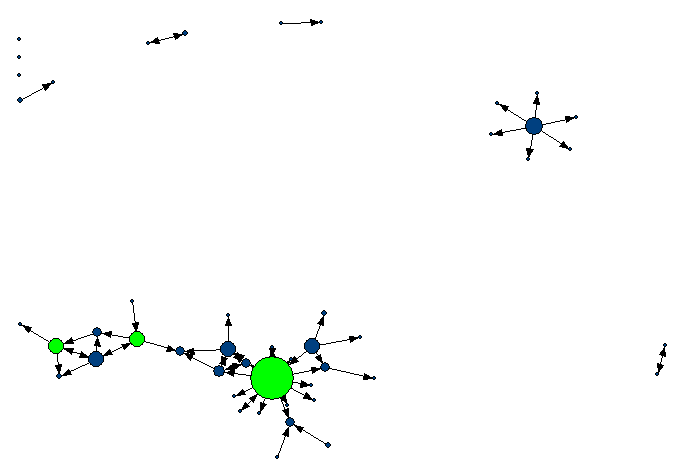}
\caption{Selfie Game with Key Players – Only Early Adopters (Seeded Developers members removed).}
\label{figure:sociogram_keyplayers}
\end{figure}

When all of the community animators are removed from the network, the 0.988 is the final fragmentation, with 0.134 as change in fragmentation (see Table 2). The final network fragmentation (Figure 7) showed that visitors were more isolated (H3) and less connected (H2) without key players, resulting in a network of small components. This result is consistent with our second hypothesis that describes animators as bridges in the network. It is also clear in Figure 7 some participants become unconnected isolates that are not socially engaged when connections to animators are removed.

\begin{figure}[!ht]
\centering 
\includegraphics[width=9cm]{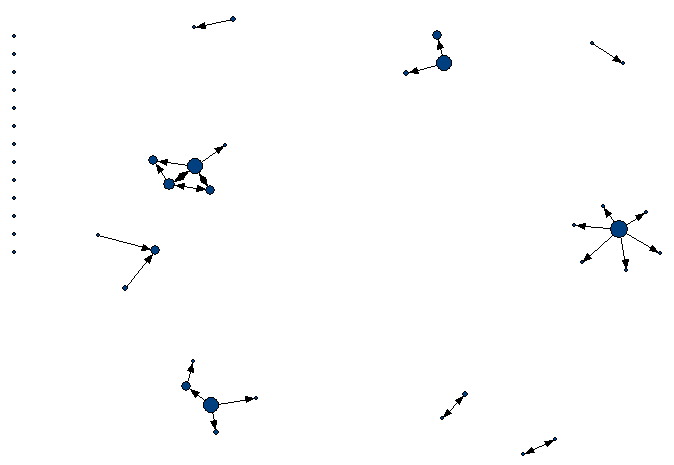}
\caption{Selfie Game Fragmentation without Any Key Players.}
\label{figure:fragmentation}
\end{figure}

\begin{table}[!ht]
\begin{flushleft}
\setlength\extrarowheight{4pt}
\centering
\small
\caption{Summary of Selfie Game Data Fragmentation. Note: High Fragmentation Means Increased Numbers of Unconnected Pairs in Network.}
\label{table:selfiegamesummary}
\begin{tabular}{m{13em}|m{1cm}m{1cm}m{1cm}}
\textbf{Selfie Game Data} & \multicolumn{3}{c}{\textbf{Fragmentation}}\\
\multicolumn{1}{c}{} & \textbf{Initial} & \textbf{Final} & \textbf{Change}\\ 
\hline
Seeded Developers (n = 2) & 0.854  & 0.947  & 0.093\\
Early Adopters (n = 3)& 0.854 & 0.895 & 0.041\\
After all key players removed & 0.854& 0.988 & 0.134 \\
\end{tabular}
\end{flushleft}
\end{table}

\section{App Data Analysis}
Among festival-goers that downloaded the app in 2015, the majority used the app for the digital program and schedule of events alone. 149 participants utilized social features. When performing KPP-NEG ten key players were identified that held potential to fragment 99\%\ of the network. KPP-POS identified 10 key players that reached 96\%\ of the network. Of the actors in the network with the highest degree of centrality, the top five were seeded developers (who obtained the most interactions while staffing the booth at the festival), and the next three were early adopters of the app. A comparison of KPP-NEG, KPP-POS, and actors with the highest degree centrality are displayed in Table 3 in terms of the distribution of seeded developers and early adopters. Further examination of the impact of community animators in encouraging the use of social features of the app is described in an analysis of fragmentation based on KPP-NEG in the proceeding section.

\begin{table}[!ht]
\setlength\extrarowheight{3pt}
\centering
\caption{Seeded Developers and Early Adopters Identified. Note: High Fragmentation Means Increased Numbers of Unconnected Pairs in Network.}
\label{table:demographics}
\begin{tabular}{p{0.2\textwidth}|p{0.04\textwidth}|p{.04\textwidth}|p{.09\textwidth}}
\hline
\textbf{} & \textbf{KPP-NEG} & \textbf{KPP-POS} & \textbf{Highest Degree Centrality}\\
\hline
Seeded Developers & 6  & 2 & 5 \\ 
\hline
Early Adopters & 4  & 8 & 5 \\ 
\hline
\end{tabular}
\end{table}

KPP-NEG effectively identified the six seeded developers that participated in recruitment at the arts festival booth (H1). In addition, four early adopters were identified with this procedure. As was the case with the Selfie Game, the early adopters identified using KPP-NEG were different than those identified using KPP-POS, with the exception of one early adopter identified in both. Only two of the seeded developers were identified using KPP-POS. Impact of each group was further examined in relation to KPP-NEG.

The network of in-app interactions before key players were removed is shown in Figure 8, the initial fragmentation in the network is 0.872.

\begin{figure}[!ht]
\centering 
\includegraphics[width=9cm]{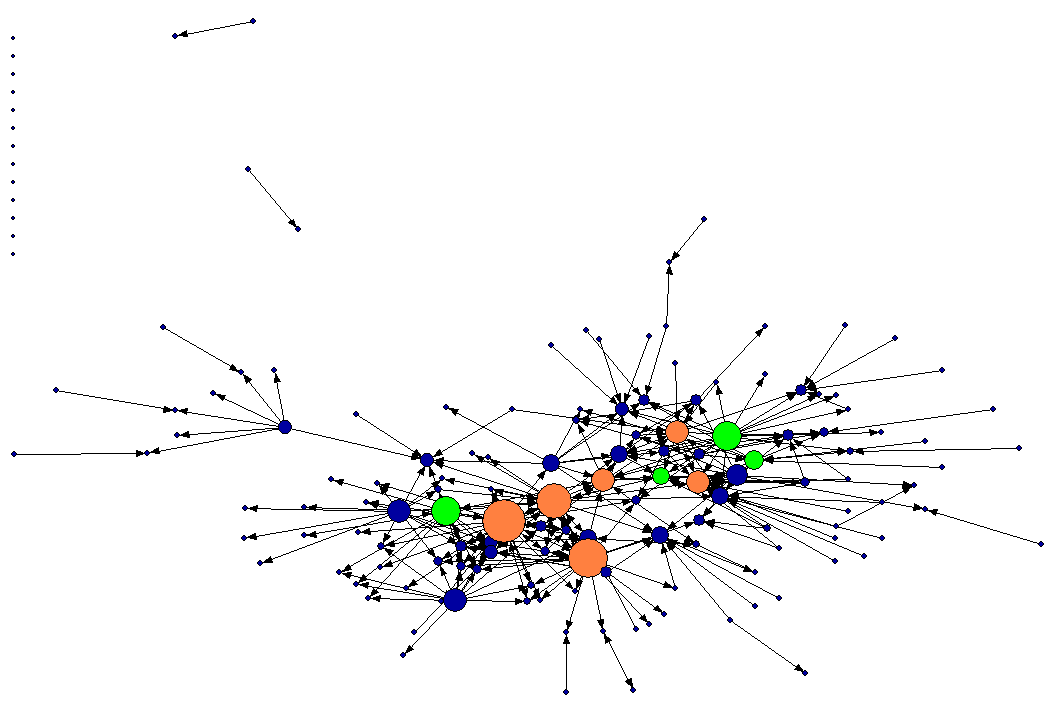}
\caption{App Data KPP-POS highlighted (Orange=Seeded Developers; Green=Early Adopters; Nodes sized by Degree Centrality).}
\label{figure:sociogram_kpppos}
\end{figure}

After removing seeded developers, fragmentation dropped to 0.986, a change of 0.114 (see Table 4). When early adopters were removed from the network, fragmentation was impacted to a lesser extent, 0.936 with a 0.064 change (see Table 4). Again, seeded developers impact on the network was roughly double that of early adopters.

\begin{figure}[!ht]
\centering 
\includegraphics[width=9cm]{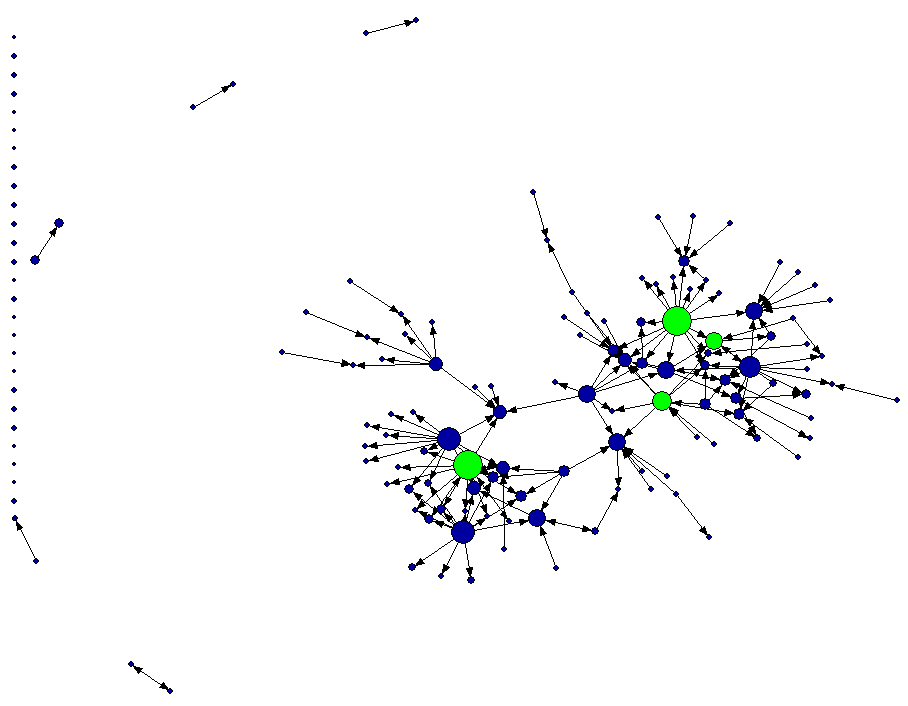}
\caption{SApp Data with Key Players – Only Early Adopters(Seeded Developers Removed).}
\label{figure:early}
\end{figure}

When all community animators were removed from the network (as illustrated in Figure 10), final fragmentation is 0.994, a 0.122 change in fragmentation (see Table 4). Comparing the visualizations of with and without community animators (Figure 8 and 10), we can see clear differences in the network in terms of overall density and cohesion (H2). When comparing the initial and final networks of the selfie game, the removal of animators had a greater impact on the network, indicating that the number of KPP-NEG could be increased depending on goal of the analysis (H3).

\begin{figure}[!ht]
\centering 
\includegraphics[width=9cm]{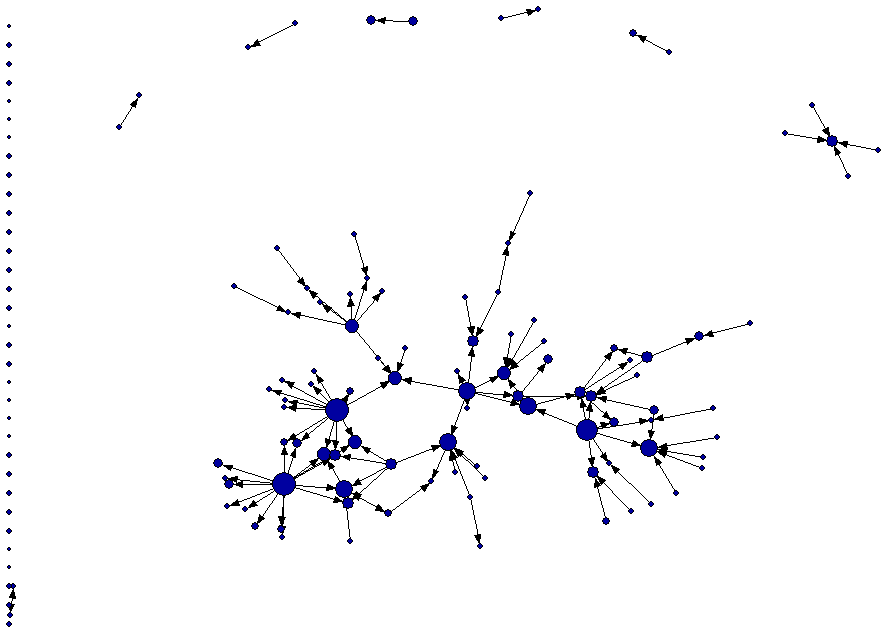}
\caption{App Data without Any Key Players.}
\label{figure:nokeyplayers}
\end{figure}

\begin{table}[!ht]
\begin{flushleft}
\setlength\extrarowheight{4pt}
\centering
\small
\caption{Summary of App Data Fragmentation. Note: High Fragmentation Means Increased Numbers of Unconnected Pairs in Network.}
\label{table:selfiegamesummary}
\begin{tabular}{m{13em}|m{1cm}m{1cm}m{1cm}}
\textbf{Selfie Game Data} & \multicolumn{3}{c}{\textbf{Fragmentation}}\\
\multicolumn{1}{c}{} & \textbf{Initial} & \textbf{Final} & \textbf{Change} \\
\hline
Seeded Developers (n = 6) & 0.872  & 0.986  & 0.114\\
Early Adopters (n = 4)& 0.872 & 0.936 & 0.064\\
After all key players removed (n = 10) & 0.872 & 0.994 & 0.122 \\
\end{tabular}
\end{flushleft}
\end{table}

\section{Discussion}
Using data collected from two socio-technical activities, we examined the network of interactions situated in the context of an arts festival promoted by a local community. Specifically, we used key player analysis of a social network analysis to operationalize, identify, and measure the impact of community animators. We explored a method of identifying community animators using algorithms to identify individuals that build cohesion among subgroups and clusters of participants. Several important results emerged.

First, we found that through key player analysis, we can identify a group of people that worked as social intermediaries to help build bridges (KPP-NEG) and expand reach of connectivity to disparate portions of the network (KPP-POS) within the Arts Festival community (RQ1). This finding was consistent with our first hypothesis, however, when comparing the use of KPP-NEG to KPP-POS, the use of KPP-NEG was more useful in identifying known animators—our own development team. The use of KPP-NEG was also useful in the ability to identify individuals that act as bridges among components in the network. Although we believe that individuals identified using KPP-NEG more accurately fit the description of Community Animators and was more successful in identifying individuals in the network that worked to animate activities in the app, the results of KPP-POS may still prove useful to researchers interested in other aspects of technology adoption. For example, if we had the explicit intention to spread a message to all users of the app or would like to survey participants with non-overlapping experiences, identifying key players using KPP-POS would provide an opportunity to focus efforts on individuals in advantageous structural positions to local clusters throughout the network.

Second, it is clear from the results of the comparison of key player analyses and the difference in fragmentation, that seeded developers’ social behavior and role can be characterized using these analyses. Although when all key players (developers and early adopters) were removed from the network of interaction, we obtained a higher fragmentation that indicates the nodes were more isolated and less dense (Figure 7 and 10). According to data presented in Tables 2 and 4, seeded developers are key players and had impact functioning as seeded animators in the network of interactions. The impact of developers doubled that of early adopters, which confirms the importance of these seeded animators in a social space to promote technology adoption. This change of fragmentation indicates that the role of community animators is important as ``bridges" \cite{borgatti2005centrality} [5] connections to begin animating participants into the socio-technical environment otherwise the participants would be disconnected components of the network (H2). This result is consistent with early adopter marketing strategies \cite{whittle2010voiceyourview}. From a community organization point of view, it is important to know who facilitate or broadcast information. These ``bridges" have the potential of becoming future partners or events promoters.

Third, following our goal of investigating seeded animator’s impact of a network of interactions; we aim to better understand community animator's impact on adoption and use of technology. Ideally, these interactions should occur among participants alone. However, based on our findings, we verified that the role of community animators is important to initiate community interactions in a social space and aid in technology adoption by informing the festival participants about a type of technology and helping them to be excited about using it (H3). Given the additional fragmentation of the network into multiple components after the removal of community animators from the selfie game, there is some indication that the role of animators played a significant role in this type of interaction as compared to lightweight interactions within the app.

\subsection{Limitation and Future Research}
Our study has some limitations worth pointing out. First, we understand that our study is a case study that may not be generalizable to other community contexts and socio-technical systems. While this study focuses on using Key Player Analysis as a method to detect community animators, we also recognize the value of mixed methods approaches to understand socio-technical systems and believe that this work would benefit from a qualitative understanding. A complete analysis of the data can include qualitative analysis to consider attitude, beliefs, and behaviors of key players and other festival participants as parameters.

Future research on this analysis will apply timestamps to examine cascading effects of seeding the interactions. We want to investigate how the impact of community animator's influence in the network after the community animator’s initiation. In addition, we see potential to test our hypothesis in different local events along the years. We want to identify and compare key players and observe the change in the degree of fragmentation. We may also consider the relative advantages of other graph invariant such as density, average tie strength, independent paths, maximum tie flow, cliques
per node, or other measures of graph cohesion to explore this or other scenarios in a different perspective. In future work for socio-technical activities, our team is working in the official mobile application for this year local community arts festival. In this new prototype, we plan to build a network of interaction through social media. We will observe how participants interact using Twitter. Participants will be the nodes in the network. Moreover, photo sharing, comments, posts and, use of the festival \textit{hashtag} will be the links.

Based on these results, we propose the identification of community animators as a proposed method to build strategy for adoption of new technology, to stage interventions within early adoption of a tool, or to identify early adopters as interview subjects. Identifying early adopters that bridge portions of the network or whom reach disparate portions of the network in a network of interaction is not obvious, especially in the context of in-app interactions of mobile technology wherein users do not wear a clear device or badge. The work of community animators is also not always visible and thus, not easily supported. Using this toolkit, strategies may be developed to facilitate technology adoption and be described in an identifiable and quantifiable way. Besides, understanding the impact of community animators is a useful strategy to identify possible interviews or user study participants. Identifying community animators (early adopters) to participate in further study evaluation could be considered as one of the direct benefits of this type of analysis.
\section{Conclusion}
This work examines the role of community animators in technology adoption. The study analyzed data collected from two socio-technical activities: a selfie game and a mobile application. A social network of interactions was constructed, and KPP-NEG and KPP-POS algorithms were used to detect individuals that build cohesion among subgroups and clusters of participants in a local event. Key player analysis was used to identify community animators and to explore the differences between roles and positions of Seeded Developers and Early Adopters. Our findings show that is possible to find community animator’s among the festival visitors through key player analysis. Further, we observed the role of community animators are critical in the building of ``bridges" among the participants (KPP-NEG). Animators were active in spreading information throughout the network, supporting technology adoption (KPP-POS). Finally, we verified that seeded developers were important for initiating participant interactions. Their role as seeded aided technology adoption and strengthened connections among actors in the network.

Within the CSCW domain, most studies have focused on the definition of brokers or promoting citizen engagement. However, this work advances the current conceptual status of early adopters and proposes the use of a social network analysis method to articulate the definition of community animators. We propose a method to quantify and measure the role of seeded animators in social interactions. This case study examined how community animators can promote interaction and create bridges among divergent participants. Finally, this work contributes to as a proposal of using this method to consider other usability studies and methods of adoption for socio-technical systems.

\section{Acknowledgements}
This work would not have been possible without the dedication of our development team, the early adopters that enthusiastically participated in our studies, and the Arts Festival coordinators that have supported this work. 

\bibliographystyle{SIGCHI-Reference-Format}
\bibliography{Community2017}

\end{document}